\documentclass[preprint,byrevtex,superscriptaddress,showpacs, prb]{revtex4}
\usepackage{graphicx}
\usepackage{dcolumn}
\usepackage{bm}
\usepackage{epsfig}
\usepackage{times}
\usepackage{amsmath}
\usepackage{amssymb}
\usepackage{textcomp}
\bibstyle{apsrev}

\begin{document}
\title{Harmonic vibrational excitations in graded elastic networks: transition from phonons to gradons}
\author{J. J. Xiao}
\affiliation{Department of Physics, The Chinese University of Hong Kong, Shatin, New Territories, Hong Kong,
China}
\author{K. Yakubo}
\affiliation{Department of Applied Physics, Graduate School of
Engineering, Hokkaido University, N13-W8, Sapporo 060-8628 Japan.}
\author{K. W. Yu}
\email{kwyu@phy.cuhk.eu.hk}
\affiliation{Department of Physics,
The Chinese University of Hong Kong, Shatin, New Territories, Hong
Kong, China}
\affiliation{Institute of Theoretical Physics, The
Chinese University of Hong Kong, Shatin, New Territories, Hong
Kong, China}

\begin{abstract}
We have identified a new type of transition from extended to localized vibrational states in one-dimensional
graded elastic chains of coupled harmonic oscillators, in which the vibrating masses or nearest-coupling force
constants vary linearly along the chain. We found that the delocalization transition occurs at the maximum
frequency of the corresponding homogeneous chain, which is in a continuous single band. Although each state in
the localized phase, called gradon, can be regarded as an impurity localized mode, the localization profile is
clearly distinct from usual impurity modes or the Anderson localized modes. We also argue how gradons may affect
the macroscopic properties of graded systems. Our results can provide insights into many analogous systems with
graded characters.
\end{abstract}
\date{\today}
\maketitle

\section{Introduction}
Vibrational normal modes and their localization properties in inhomogeneous or disordered media have attracted
much interest. \cite{Lieb68, Wallis68, Yip84, Tamura88, Yakubo03, Chan04} Notable examples include
disorder-induced localized states (i.e., the Anderson localization of lattice vibrations), fractal-induced
localization (i.e., fractons \cite{YakuboRMP}), modes contributing the boson peak, \cite{Kantelhardt01} the
connection of excess of low-frequency modes with the jamming-unjamming transition in particle packings,
\cite{Silbert05} intrinsic localized modes (ILMs) caused by nonlinearity and/or discreteness in systems,
\cite{Campbell04} and the well-known defect modes in crystals including impurities \cite{Wallis68,Defect} etc.
Surface phonon modes in semiconductors are also typical localized vibrational excitations in which atomic
displacements decrease exponentially into the bulk. \cite{Fritsch99} These localized vibrational modes strongly
affect properties of condensed matter. It is crucial to reveal characteristics of localized modes for
understanding thermal, elastic, and acoustic properties and/or electron-phonon interactions in materials.

From the viewpoint of localization mechanisms, there exist two types of localized modes in
noninteracting systems. The first type is a consequence of interference of coherent vibrational
waves due to diffusive scattering. Localized modes of this type include the Anderson localization
of classical waves \cite{Stephen83,PSheng95} and fractons in fractal systems. \cite{YakuboRMP} In
these cases, localization occurs whenever the elastic mean free path becomes comparable to the
inverse wave vector of the wave (the Ioffe-Regel condition). In particular, the Anderson
localization of elastic waves in disordered solids has been extensively studied theoretically and
experimentally. \cite{Elliott03,Feddersen91,Foret96} Now, we have obtained vast amount of knowledge
on the Anderson localization and the Anderson transition, which involve similarities to and
differences from quantum particle localization. \cite{Brandes04} The mode pattern of an Anderson
localized state possesses an exponential tail and the position of the localization center is
unpredictable. The scaling theory \cite{Abrahams79} shows all vibrational modes are localized in
one- and two-dimensional disordered systems while one has the localization-delocalization
transition in three dimensions, which is the same with the Anderson transition of quantum particles
in the orthogonal universality class. \cite{Mehta91} The delocalization transition occurs in a
continuous band without any singularity of the spectrum at the transition frequency
$\omega_{\text{c}}$. The localization length $\xi$ diverges as a power law of
$|\omega-\omega_{\text{c}}|$. The other type of localizations is due to confinement by impurities
such as defect modes, surface modes, or ILMs. \cite{Maradudin71} This type of localized modes is
rather trapped by local potentials. The mode pattern of such an impurity localized mode has also an
exponential tail and the position of the localization center is located at the impurity site. For
instance, the light-mass impurity mode of a one-dimensional elastic chain has the mode pattern of
$u_{n}\propto [-m_{0}/(2m-m_{0})]^{|n|}$, where $m$ and $m_{0}$($<m$) are the host and impurity
masses, respectively, and $n$ is the site number ($n=0$ represents the impurity site). The
frequency of this light-mass impurity mode is given by
$\omega_{\text{imp}}^{2}=m^{2}\omega_{\text{max}}^{2}/m_{0}(2m-m_{0})$, where $\omega_{\text{max}}$
is the maximum frequency of the homogeneous chain with mass $m$ for all sites. Since $m_{0}<m$,
$\omega_{\text{imp}}$ is always larger than $\omega_{\text{max}}$, which implies that the
light-mass impurity localized mode has an isolated spectrum outside of the band. This is considered
as a common feature of impurity localized modes. Therefore, it seems that no
localization-delocalization transition induced by impurities occurs in a single band. This is,
however, not obvious. If a delocalization transition in a continuous spectrum is realized by
impurity modes, properties of localized vibrations and the transition would be quite different from
those for conventional localized modes and transitions.

In this work, we consider vibrational problems of a graded elastic systems which are modelled by
graded spring-mass networks. Studying one-dimensional (1D) systems with graded masses and graded
force constants, we show a new type of transition from extended to impurity localized modes. Some
characteristic properties of these localized modes are also elucidated. Graded elastic systems can
be found everywhere in nature. Mass density of an elastic body subject to a gravitational field is
low near the top of the object but high near the bottom, such as sediment at the bottom of ocean,
which can be regarded as an example of graded elastic systems. Similar localized modes to our
results can be expected in it. A carbon nanotube which can be regarded as folding a graphite sheet
into a tube is another example of graded elastic materials. \cite{Mele05} The force constant can be
stronger at the central part of the tube but weaker at both ends. Thus there is a gradient of force
constant along the nanotube. Particularly, by virtue of the analogy of the vibrational problem with
many other problems, such as electromagnetic wave propagation \cite{Nakayama92,Nakayama97} in which
the index of refraction plays the same role as the mass, one would expect similar localizations in
other type of graded structures such as quantum or photonic systems. It is possible to apply our
model to graded plasmonic nanoparticle chains which may find prominent applications in nanooptics.
In fact, there have been many efforts directed to graded systems in both artificial composites
\cite{DongPRB03} and materials in nature (e.g., biological cells \cite{HuangPRE03}). One further
notices that in contrast to a simple sharp interface, it has been studied that the presence of
gradient in a interface leads to frequency-dependent properties. \cite{Vollmann04, Berezovski00}
Although we treated 1D chains in this work, our findings form the basis for higher dimensional
networks in which results are more complicated and intriguing and we will present them elsewhere.
\cite{xiao}

In what follows, we first briefly describe, in Sec.~II, the graded elastic network model, and establish the
formalism for studying eigenmodes. In Sec.~\ref{sec:gradmass}, we present our results mainly in the graded mass
model and similar results are in the graded force constant model . Finally conclusions and discussions on
our results are given.

\section{model and formalism}
\begin{figure}[hptd]
  \includegraphics[scale=0.5]{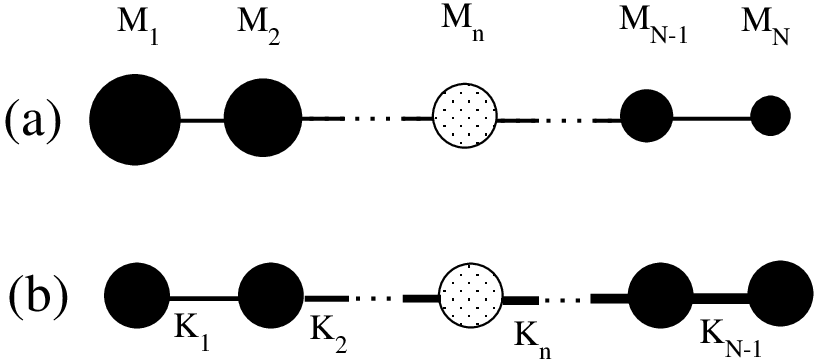}
  \caption{/Xiao, Yakubo, and Yu}
  \label{fig:fig0}
  \end{figure}

The elastic vibrational problem of a uniform three-dimensional system is simply modelled by a cubic lattice of
harmonic oscillators where different components of the displacement decouple. \cite{Kantelhardt01} However, when
there exists a uniaxial gradient of coupling constants (or
masses), the equation of motion is only partially
separable. Assuming that the network is periodic in the directions perpendicular to the gradient direction
(e.g., $z$-axis), one reaches an effective one-dimensional problem. \cite{xiao} Without loss of generality, let
us start with a one-dimensional spring-mass model as depicted in Fig.~\ref{fig:fig0}. The mass $M_n$ and the
spring constant $K_n$ connecting to the right nearest-neighbor are labelled sequently with $n=1, 2, \cdots, N$.
The motion of each mass is governed by the following equations \cite{Lieb68}
\begin{equation}
\label{eq:motion}
M_n \ddot
u_n=K_{n-1}(u_{n-1}-u_{n})+K_{n}(u_{n+1}-u_n),
\end{equation}
where $u_n$ denotes the scalar displacement of mass $M_n$ at the
site $n$, and the derivative $\ddot u_n$ is taken with respect to
time $t$. We omit damping and further couplings to avoid
complications, the inclusion of which do not change the results
qualitatively. Decomposing the amplitude $u_{n}$ by the
eigenvectors $\boldsymbol{e}(\lambda)$ of the mode $\lambda$ as
$u_n = \sum_{\lambda}Q_{\lambda}(t)e_{n}(\lambda)$, where
$Q_{\lambda}(t)$ is a time-dependent expansion coefficient which
behaves as $Q_{\lambda}(t) \sim \exp(-i\omega_{\lambda}t)$, we
simplify Eq.~(\ref{eq:motion}) to the eigensystem of
\begin{equation}
\label{eq:matrix}
-\omega^2  \boldsymbol{e}= \boldsymbol{T} \cdot
\boldsymbol{e},
\end{equation}
where the matrix $\boldsymbol{T}$ characterizes the mass-weighted coupling interactions of the neighboring
masses connected by the springs that form a lattice with lattice
constant $a=1$. The system size $L$ is then
given by $aN$. Assuming free boundary conditions, the $N\times N$ matrix $\boldsymbol{T}$ becomes tridiagonal
and with elements
\begin{widetext}
\begin{eqnarray}
T_{mn}=-\Big[\frac{(1-\delta_{n1})K_{n-1}}{M_{m}}+\frac{(1-\delta_{nN})K_{n}}{M_{m}}\Big]\delta_{mn}
+\frac{K_{n}}{M_{m}}(\delta_{mn-1}+\delta_{mn+1}),
\end{eqnarray}
\end{widetext}
where $\delta$ is the Kronecker delta. In this regard, further
incorporating the next-nearest coupling force constant
$K_n^{\text{next}}$ results a quintuple matrix $\boldsymbol{T}$.
We have also attempted periodic boundary conditions, i.e., let
$M_n$ connected to $M_1$ by $K_n$, very similar results are
observed. The nontrivial $N$ solutions of $\boldsymbol{e}$ and
$\omega^{2}$ in Eq.~(\ref{eq:matrix}) are thus determined by
diagonalizing $\boldsymbol{T}$.

In this paper, we will consider two different cases in a one-dimensional network:\\ (1) The graded mass case
with
\begin{equation}
\label{eq:gradmass}
M_n=M_0-C_{\text{M}} n/N,
\end{equation}
and fixed $K$ [see Fig.~\ref{fig:fig0}(a)]. In this case, the node mass decreases linearly from the left end
site $n=1$ to $n=N$.\\
(2) The graded force constant case with
\begin{equation}
\label{eq:gradforce}
K_n=K_0+ C_{\text{K}} n/N
\end{equation}
and fixed $M$ [see Fig.~\ref{fig:fig0}(b)]. Here, $M_0$ and $K_0$ are the maximum mass and the
minimum force constant, and $C_{\text{M}}$ and $C_{\text{K}}$ are the gradient coefficients in the
two models, respectively. Note that the conditions of $C_{\text{M}}=0$ and $C_{\text{K}}=0$ recover
the homogeneous chain. It is also emphasized that our graded models can be easily extended to
infinite $N$ (namely, infinite system size $L$). The infinite graded chain does not correspond to
the continuum graded chain, but to the infinitesimal gradient limit.

\section{Numerical Results}
\label{sec:gradmass} 
\begin{figure}[hptd]
  \includegraphics[scale=0.5]{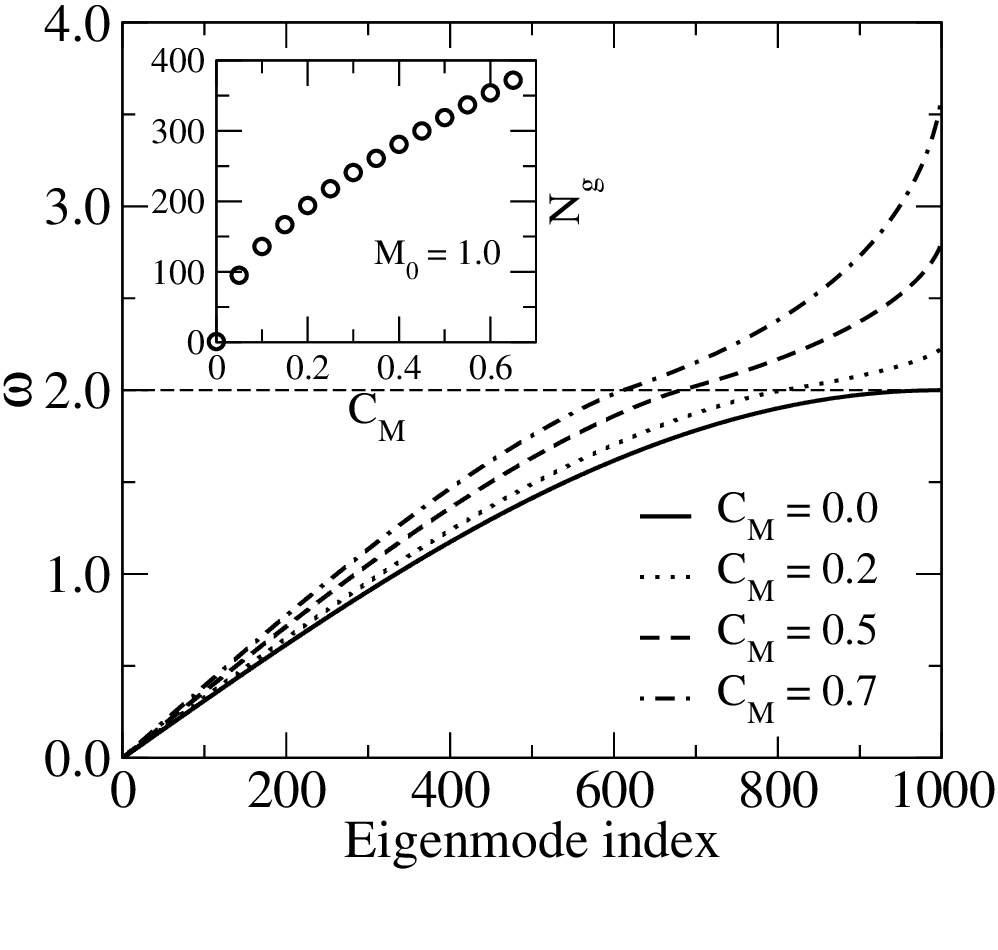}
  \caption{/Xiao, Yakubo, and Yu}
  \label{fig:m1}
  \end{figure}

\begin{figure}[hptd]
  \includegraphics[scale=0.5]{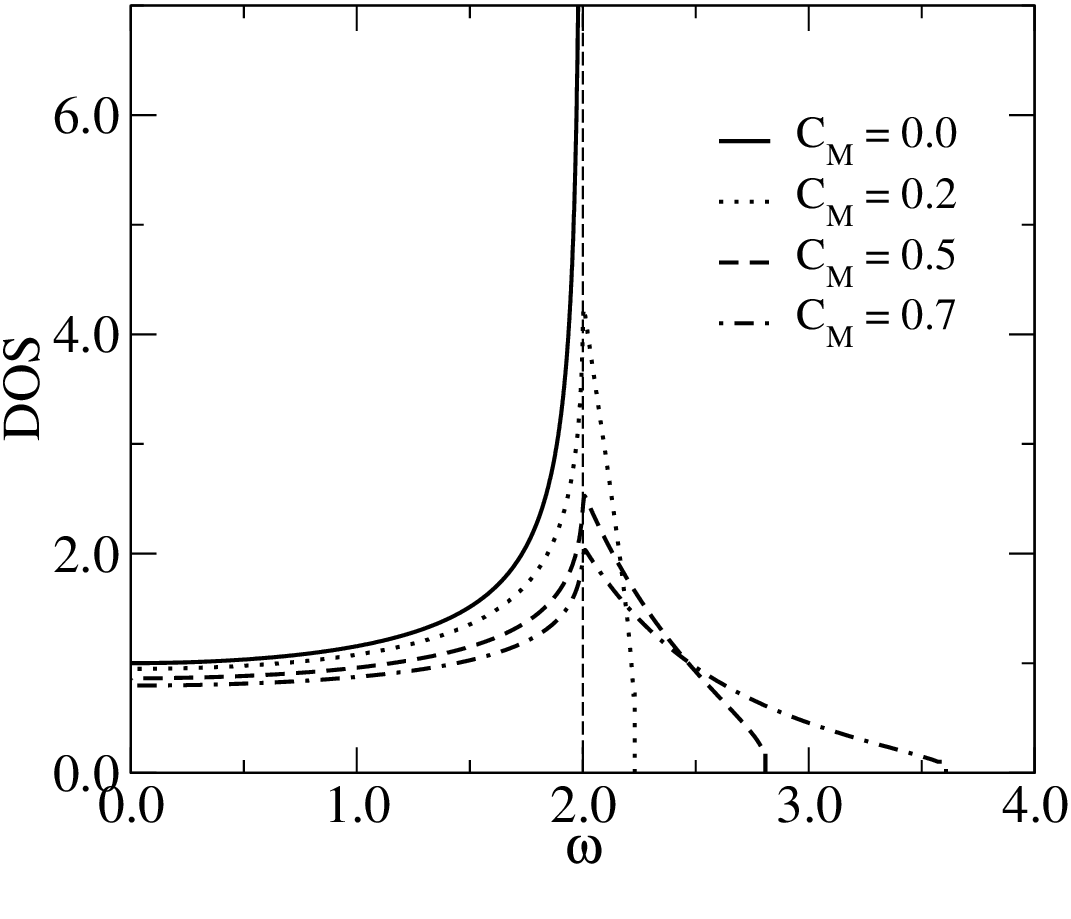}
  \caption{/Xiao, Yakubo, and Yu}
  \label{fig:dos}
  \end{figure}

\begin{figure}[hptd]
  \includegraphics[scale=0.8]{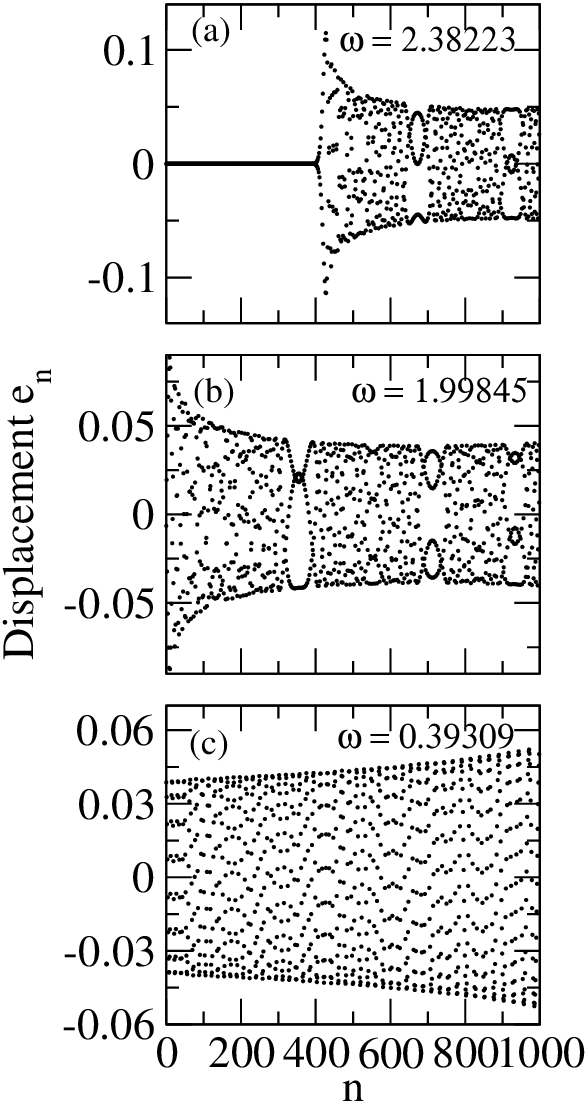}
  \caption{/Xiao, Yakubo, and Yu}
  \label{fig:m2}
  \end{figure}

In this section, we mainly concentrate on the graded mass model with $N=1000$. A further
increase in $N$ does not change our results (except for Fig.~\ref{fig:ck} as shown later). The dispersion
relation of the vibrational modes in a crystal gives rich
information about properties of phonons. In the
graded elastic network, however, the wavenumber is not conserved because of the lack of translational symmetry
in the system. This does not allow us to consider the dispersion relation for our model. Nevertheless, the
relation between the eigenfrequency $\omega_{\lambda}$ and the mode index $\lambda$ in ascending order with
respect to $\omega_{\lambda}$, which can be regarded as a pseudo-dispersion relation, helps us to interpret
vibrational modes in the graded system. Figure~\ref{fig:m1} shows the eigenfrequency $\omega_{\lambda}$ as a
function of the eigenmode index $\lambda$ for gradient coefficients $C_{\text{M}}=0.0$, $0.2$, $0.5$, and $0.7$.
$M_0=1$, $K_0=1$ and $C_{\text{K}}=0$ are used. Note that the relation for the homogeneous elastic chain
($C_{\text{M}}=0.0$) represents the usual dispersion relation. It is clear that the presence of the gradient
alters the (pseudo-) dispersion relation, particularly beyond the frequency
$\omega_{\text{c}}=2\sqrt{K_0/M_0}=2.0$ (dashed horizontal line) which is the maximum eigenfrequency of the
homogeneous elastic chain with $M_{n}=M_{0}$ for any $n$ (the solid curve). The number $N_{\text{g}}$ of
high-frequency excitations with $\omega
>\omega_{\text{c}}$ depends on the gradient coefficient
$C_{\text{M}}$ as shown by the insert of Fig.~\ref{fig:m1}. The maximum frequency of the graded
chain is given by $\omega_{\text{max}}=2\sqrt{K_{0}/M_{N}}$ ($2\sqrt{K_{N}/M_{0}}$) in the graded
mass (force constant) model. Let us call these high energy excitations ``gradons" which are clearly
distinct from low frequency excitations as we will show later. The fact that $\omega_{\text{c}}$ is
a characteristic frequency in the graded elastic chain is more explicitly found in the density of
states (DOS) defined by
\begin{equation}
D(\omega)=\frac{1}{N}\sum_{\lambda}\delta(\omega-\omega_{\lambda}).
\end{equation}
The DOS shown in Fig.~\ref{fig:dos} clearly exhibits a singular maximum peak at
$\omega_{\text{c}}=2.0$ for $C_{\text{M}}>0$. This behavior persists for various gradient
coefficients $C_{\text{M}}$. The low-frequency DOS is in accordance with the usual asymptotic Debye
density of vibrational states,  both in the homogeneous case and the graded case. From the DOS, we
expect that modes with $\omega>\omega_{\text{c}}$ are qualitatively different from modes with
$\omega<\omega_{\text{c}}$. In order to clarify this point, we plot the mode patterns
$\boldsymbol{e}(\lambda)$ in Fig.~\ref{fig:m2}, where three typical modes with
$\omega=2.38223(>\omega_{\text{c}})$, $1.99845(\approx \omega_{\text{c}})$, and
$0.39309(<\omega_{\text{c}})$ are presented. These mode patterns show that modes with
$\omega>\omega_{\text{c}}$ (namely, gradons) are confined at the ``light" mass side, whereas those
with $\omega<\omega_{\text{c}}$ (phonons) are extended. As will be shown later, these confined
modes are qualitatively different from conventional localized modes. Nevertheless, we use the word
{\it localized modes} for these confined states, because the present model exhibits a drastic
change in spatial extent of modes. One can find a beating pattern in the wave profile [see
Fig.~\ref{fig:m2}(b)], which implies that the mode is composed of plural wavenumber components. The
degree of mixing of different wavenumber components can be quantified by the full width at half
maximum (FWHM) of the Fourier amplitude $|F_j(\boldsymbol{e})|=|\pi\sum_{n=1}^N e_n \exp{[2\pi i
(j-1)(n-1)/N]} /\sqrt{N}|$. Indeed, the Fourier transform of the mode pattern in
Fig.~\ref{fig:m2}(a) has a finite width around the average wavenumber as shown in Fig.~5(a). Figure
5(b) gives the FWHM as a function of $\omega$ for several values of $C_{\text{M}}$. The FWHM shows
singularity at $\omega=\omega_{\text{c}}$ as in the case of the DOS. For well-extended states for
$\omega\ll \omega_{\text{c}}$, the FWHM must be very small, which implies that we can approximately
define the Bloch wavenumber $k$ for modes in this very low frequency regime. It is then possible to
regard Fig.~\ref{fig:m1} as a usual dispersion relation between $\omega$ and $k$ for $\omega \ll
\omega_{\text{c}}$.

In order to quantify the degree of localization of excited modes, we consider the inverse
participation ratio \cite{Toulous79,Wegner80} (IPR),
\begin{equation}
\label{eq:IPR}
P^{-1}=\frac{\sum\limits_{n=1}^Ne_n^4}{\Big(\sum\limits_{n=1}^N
e_n^2\Big)^2}.
\end{equation}
The IPR defined by Eq.~(\ref{eq:IPR}) gives a measure of the number of sites contributing
significantly to a given eigenstate. If the state is strongly localized, $P^{-1}$ becomes
proportional to $1/\xi^{d}$, where $\xi$ is the localization length, while for a fairly extended
state it is proportional to $1/L^{d}$ because all sites contribute to the state. We show in Fig.~6
the IPR of vibrational modes in the graded mass model as a function of the eigenfrequency for
various gradient coefficients $C_{\text{M}}$. At low frequencies ($\omega<\omega_{\text{c}}$),
small and almost flat IPR indicates fairly well-extended excitations across the whole network. The
IPR increases dramatically, however, as the frequency exceeds $\omega_{\text{c}}=2.0$. This shows
that modes with $\omega>\omega_{\text{c}}$ are spatially localized. These are in accord with the
excitation shapes depicted in Fig.~\ref{fig:m2}. As seen in the inset of Fig.~6, the small step of
the IPR at $\omega=\omega_{\text{c}}$ is somewhat broad. This is a consequence of the finite size
effect of our $N=1000$ system. We confirmed that the step becomes sharper as increasing $N$ (figure
not shown here).

Due to the duality between force constant and mass, the graded force constant model represented by
Eq.~(\ref{eq:gradforce}) is essentially the same with the graded mass model we discussed above.  Indeed, we
basically observed the same results as those in the graded mass model. In Fig.\ref{fig:fig7}, for example, we
present three typical excitation states with $\omega=2.27844(>\omega_{\text{c}})$, $1.99911(\approx
\omega_{\text{c}})$, and $0.377618(<\omega_{\text{c}})$ for the graded force constant model with
$C_{\text{K}}=1.0$. These mode patterns are quite similar to those in the graded mass model shown in
Fig.~\ref{fig:m2}. Further results on the pseudo-dispersion relation, the DOS, the FWHM, and the IPR show also
similar behaviors to those in the graded mass case.

From all the results presented above, we can conclude that an infinite graded elastic chain exhibits the
localization-delocalization transition at the frequency $\omega_{\text{c}}$ being the maximum eigenfrequency of
the homogeneous elastic chain with $M_{n}=M_{0}$ (or $K_{n}=K_{0}$) for any $n$. In the frequency regime
$\omega<\omega_{\text{c}}$, we have phonon-type modes extended over the whole system, while modes with
$\omega>\omega_{\text{c}}$ (namely, gradons) are localized in the lighter mass side (or harder force constant
side). It should be noted that this delocalization transition (the phonon-gradon transition) occurs in a single
band as shown in Fig.~3. The mechanism of gradon localization is considered as follows. A mode with a frequency
$\omega$ larger than $\omega_{\text{c}}$ cannot have amplitudes in the heavier (or softer) region because masses
(or force constants) in this region are too heavy (or too soft) to vibrate with the frequency $\omega$. As a
consequence, the mode pattern has finite amplitudes only in the lighter (or harder) region, while no amplitude
in the heavier (or softer) region. The boundary between these two regions is determined by
$\omega=2\sqrt{K_{n}/M_{n}}$. Therefore, the localization center site $n_{\text{c}}$ which is the position with
the maximum amplitude [see Fig.~\ref{fig:m2}(a)] is given by
\begin{equation}
\label{eq:nc}
n_{\text{c}}=
\begin{cases}
\displaystyle
\frac{NM_{0}}{C_{\text{M}}}\left(1-\frac{\omega_{\text{c}}^{2}}{\omega^{2}}\right)
& \text{graded mass model}, \\
&\\
\displaystyle
\frac{NK_{0}}{C_{\text{K}}}\left(\frac{\omega^{2}}{\omega_{\text{c}}^{2}}-1\right)
& \text{graded force constant model}.
\end{cases}
\end{equation}
The fact that values of $n_{\text{c}}$ of the modes shown by
Fig.~\ref{fig:m2}(a) and Fig.~7(a) are predicted by
Eq.~(\ref{eq:nc}) as $n_{\text{c}}=423$ for Fig.~\ref{fig:m2}(a)
and $298$ for Fig.~7(a) shows the validity of our interpretation
of gradon localization. This mechanism of localization is
essentially the same with that of impurity localized modes.
Therefore, gradons belong to a kind of confined modes by
impurities, and the phonon-gradon transition in a single band is
induced by impurity localized modes.

Properties of this new type of the delocalization transition are largely different from usual transitions
occurring in a single band, such as the Anderson transition. First of all, the gradon transition can occur even
in a one-dimensional graded system, while the Anderson transition requires three or more dimensions. This is not
surprising, because the scaling theory cannot be applied to the gradon transition. Furthermore, a localized mode
in a usual sense has a finite localization length $\xi$ even if the system size is infinite. On the contrary,
the localization length of a gradon, which is defined as the size of vibrating region and approximately given by
$N-n_{\text{c}}$, becomes infinite if the system size $N$ tends to
infinity as seen from Eq.~(\ref{eq:nc}).
Gradon modes are localized only in the sense that a part of the whole system has vibrational amplitudes. The
ratio of the vibrating region to the whole system is given by $1-M_{0}(1-\gamma^{2})/C_{\text{M}}$ for the
graded mass model and $1-K_{0}(\gamma^{-2}-1)/C_{\text{K}}$ for the graded force constant model, where
$\gamma=\omega/\omega_{\text{c}}$. Although it is known that there exist one-dimensional delocalization
transitions in correlated disordered systems, \cite{Dunlap90} the gradon transition is distinguished from this
type of transitions at this point. The fact that the DOS shows the singularity at the transition frequency is
also a feature of the phonon-gradon transition, which contrasts strikingly with the Anderson transition without
any spectral singularity at the critical point. For the graded mass model, each state in the vibrational DOS
above $\omega_{\text{c}}$ corresponds to a light-mass impurity mode mentioned in Sec.~I. The linear distribution
of light-mass impurities in the graded mass model forms a continuous spectrum of impurity modes above
$\omega_{\text{c}}$. The whole spectral structure shown in Fig.~3 is constructed by connecting two different
spectra, the phonon-type excitation spectrum and the impurity mode one above $\omega_{\text{c}}$.

As seen from Figs.~\ref{fig:m2} and \ref{fig:fig7}, mode pattern profiles of gradon excitations are highly
asymmetric due to the asymmetricity of the graded elastic network. A gradon mode in the heavier (or softer) side
of the localization center $n_{\text{c}}$ has an exponential tail with a large damping factor $d_{0}^{-1}$ as
$\sim\exp[-(n_{\text{c}}-n)a/d_{0}]$. The width of the localization front (i.e., the decay length) $d_{0}$ seems
to be independent of $\omega$. On the contrary, in the lighter (or harder) side of $n_{\text{c}}$, vibrational
amplitudes decrease very slowly (almost constant) as increasing $n-n_{\text{c}}$. We should remark that the
gradon transition remains sharp even for a finite system. This is because the width $d_{0}$ is much smaller than
the system size $L$. This also contrasts to a conventional transition described by the scaling theory, in which
the finite system size reduces the transition to a broad crossover.

\section{conclusion and discussion}
In summary, we identified a type of transition from extended to localized states in the
one-dimensional graded elastic network described by Eq.~(\ref{eq:motion}) with
Eq.~(\ref{eq:gradmass}) or (\ref{eq:gradforce}). The delocalization transition occurs at
$\omega=\omega_{\text{c}}$, where $\omega_{\text{c}}$ is the maximum frequency of the homogeneous
network with $M_{n}=M_{0}$ (or $K_{n}=K_{0}$) for any $n$. The transition is totally distinguished
from conventional disorder induced delocalization transition such as the Anderson transition in the
following sense: (1) The delocalization transition occurs even in a one-dimensional system in
contrast to the Anderson transition requiring three or more dimensions. (2) Localized vibrational
states which is called gradons in the graded elastic network are classified into impurity localized
modes though the spectral density of states (DOS) of gradon modes is continuous. (3) The
localization length of a gradon becomes infinite if the system size tends to infinity. (4) At the
transition frequency $\omega_{\text{c}}$, the DOS shows a singularity, which is a consequence of
the connection of the phonon-type excitation spectrum to the one of impurity mode  at
$\omega=\omega_{\text{c}}$. The DOS of infinite graded chain can in fact be expressed as an average
of the DOS of homogeneous chains over a range of masses. (5) The mode pattern of a gradon
excitation is highly asymmetric around the localization center which is the position with the
maximum amplitude. (6) The gradon transition remains sharp even for a finite system because the
width of the localization front is much smaller than the system size.

It would be of great interests to study the general graded profiles. In this regard, further calculations with
other graded profiles have been attempted and qualitatively similar results have been obtained. For instance,
the power-law graded profile still confirms the existence of the transition from phonons to gradons. However,
the transition frequency $\omega_{\text{c}}\neq 2$ any longer; it depends on the detailed graded profile.
Higher-dimensional graded networks are more realistic models. These networks are possibly in two or three
dimensional crystals, in which the vibrational modes can be partitioned in the dispersion relations, due to the
presence of bounded surface elastic waves. \cite{xiao} In current considerations, we have not included any
damping or non-neighboring couplings and open boundary conditions are enforced. We believe the inclusion of
further coupling or damping will not change the conclusions much, although inclusion of a damping can change the
normal modes with an overall time decaying factor. In fact,
further calculations with the next-nearest coupling
$K^{\text{next}}_{n}$ in the model show enhancement of the degree
of localization, increasing of the
transition frequency $\omega_{\text{c}}$ but decreasing of the
number of localized modes. \cite{xiao} The localization tail
still exhibits an exponential rather than power-law decay, the latter of which occurs only in systems with
long-range couplings. \cite{xiao} Localized states in high
dimensions with vector displacements are also
possible, but the bending and stretching force constants can introduce a new characteristic length scale and
more complicated spectrum in such systems.

It is also natural to think about how physical quantities are affected by the existence of gradons. Since the
DOS possesses a sharp peak at the phonon-gradon transition frequency $\omega_{\text{c}}$, we could expect a
remarkable structure near $T=T_{0}\equiv \hbar\omega_{\text{c}}/K_{\text{B}}$ in the heat capacity, just as the
hump of the specific heat capacity due to boson peak. \cite{Kantelhardt01} Furthermore, we believe the
temperature dependence of the thermal conductivity $\kappa(T)$ is strongly affected by gradons due to the fact
that the diffusion constant for localized gradon modes is zero, which facilitates locking of the heat flux. It
is because if we neglect the anharmonic interaction between phonons and gradons, the thermal conductivity would
become constant for $T\gg T_{0}$. One cannot, however, rule out the anharmonic interactions between phonons and
gradons at high temperatures, and the decay processes from gradons
to phonons give an additional thermal
conductivity. As a result, the temperature dependence of
$\kappa(T)$ for $T\gg T_{0}$ would be different from
$\kappa(T)$ for crystals, which provides experimental evidence of gradon modes in the graded elastic networks.
Further incorporating the gradon results with diatomic model,
anharmonicity of nearest neighbor interaction, and
rotational degree of freedom \cite{xiao} is also rewarding. We
expect experiments be done to observe
gradons, or even its analogous like spin-wave gradons, quantum gradons, and photonic gradons etc. The present
results and those in higher dimensions \cite{xiao} will have implications in surface acoustic wave applications
as diverse as touchscreens, pizeoelectric material design, and
integrated microresonator systems etc, and also
provide benefits to a wide spectrum of problems such as earthquake study or oil exploration.

\acknowledgments{The authors would like to acknowledge useful discussion with professor T.
Nakayama. This work was supported in part by the RGC Earmarked Grant of the Hong Kong SAR
Government (K.W.Y), and in part by a Grant-in-Aid for Scientific Research from Japan Society for
the Promotion of Science (No.~16360044).}

\newpage
\begin{center}
\textbf{Figure Captions}
\end{center}

\noindent\textbf{FIG. 1}.Schematic pictures of the graded elastic networks. (a) The graded mass model, and (b)
the graded force constant model.

\hfill

\noindent\textbf{FIG. 2}. The pseudo-dispersion relation for the graded mass models with gradient coefficients
$C_{\text{M}}=0.0$, $0.2$, $0.5$, and $0.7$ and $M_{0}=1.0$. $C_{\text{M}}=0$ corresponds to the homogeneous
case [i.e., $k=n\pi/L$ and $\omega=2\omega_0\sin(ka/2)$]. Inset shows the number of gradons $N_{\text{g}}$ as a
function of $C_{\text{M}}$.

\hfill

 \noindent\textbf{FIG. 3}. The densities of states for the graded mass model with
$M_{0}=1.0$ and various $C_{\text{M}}$.

 \hfill

 \noindent\textbf{FIG. 4}.Three typical excitation states of the graded mass model with
$M_{0}=1.0$ and $C_{\text{M}}=0.7$ at (a) $\omega=2.38223(>\omega_{\text{c}})$, (b) $\omega=1.99845(\approx
\omega_{\text{c}})$, and (c) $\omega=0.39309(<\omega_{\text{c}})$.

 \hfill

 \noindent\textbf{FIG. 5}.(a) The Fourier transform of the mode pattern shown
by Fig.~4(a). (b) The full width at half maximum of Fourier transforms of mode patterns of the graded mass model
with $M_{0}=1.0$ as a function of $\omega$.

\hfill

 \noindent\textbf{FIG. 6}. The
inverse participation ratio (IPR) of modes excited in the graded mass model with $M_{0}=1.0$ plotted as
functions of the eigenfrequency. Inset shows the magnified IPR around $\omega=2.0$.

  \hfill

 \noindent\textbf{FIG. 7}. Three typical excitation states of the graded
force constant model with $K_{0}=1.0$ and $C_{\text{K}}=1.0$ at (a) $\omega=2.27844(>\omega_{\text{c}})$, (b)
$\omega=1.99911(\approx \omega_{\text{c}})$, and (c) $\omega=0.377618(<\omega_{\text{c}})$.

\newpage

\newpage

  \newpage

    \newpage

\newpage
\begin{figure}[hptd]
  \includegraphics[scale=0.6]{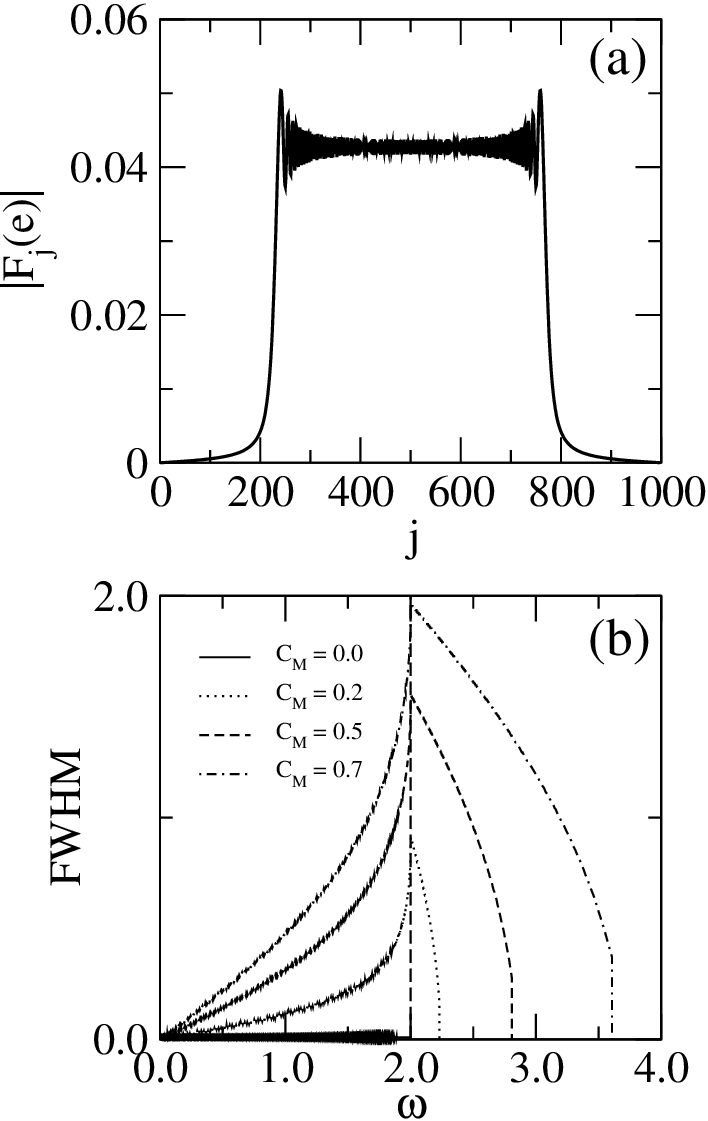}
  \caption{/Xiao, Yakubo, and Yu}
  \label{fig:m3}
  \end{figure}

  \newpage
\begin{figure}[hptd]
  \includegraphics[scale=0.5]{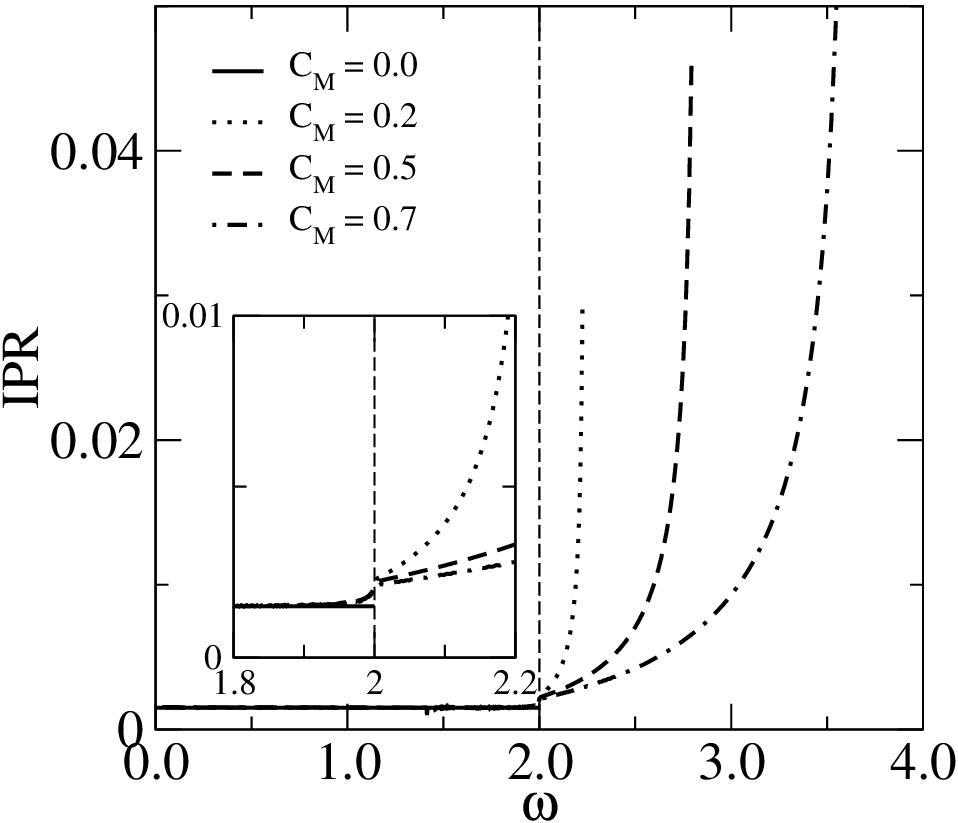}
  \caption{/Xiao, Yakubo, and Yu}
  \label{fig:ck}
  \end{figure}

  \newpage
\begin{figure}[hptd]
  \includegraphics[scale=0.8]{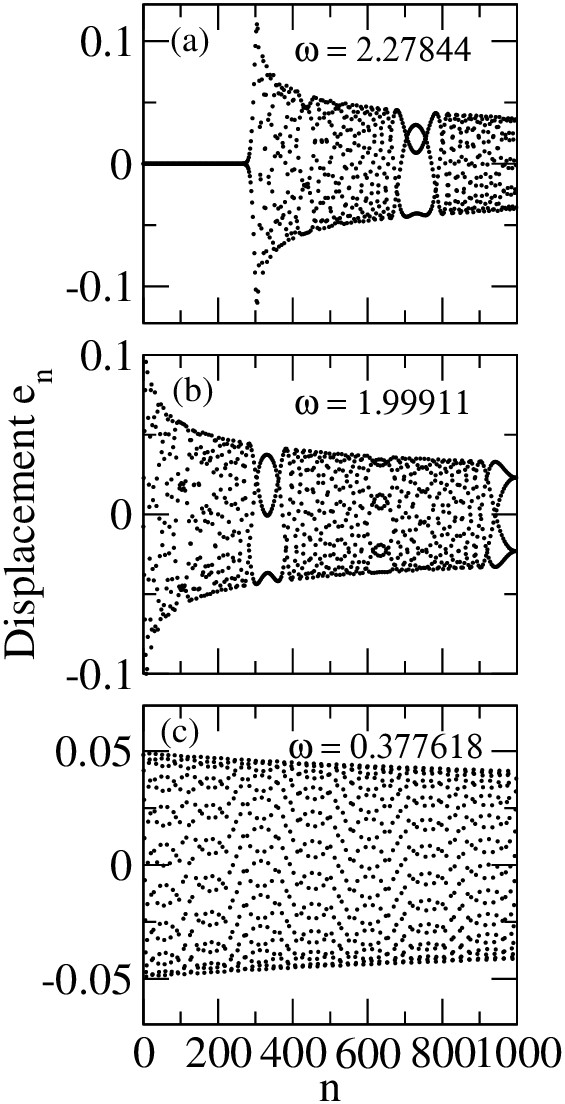}
  \caption{/Xiao, Yakubo, and Yu}
  \label{fig:fig7}
  \end{figure}

\end{document}